\documentclass[a4paper]{panl}
\usepackage{cite}
\usepackage{wrapfig}
\usepackage{graphicx}
\usepackage{amssymb}
\usepackage{amsfonts}
\usepackage{amsmath}
\usepackage{longtable}
\usepackage{rotating}
\usepackage{lscape}
\usepackage{epsfig}
\usepackage{multirow}

\originalTeX

\begin{document}

\title{A New Mechanism for Sympathetic Cooling of Atoms and Ions in Atomic and Ion-Atomic Traps}
\maketitle
\authors{V.~S.~Melezhik\/$^{a,b,}$\footnote{E-mail: melezhik@theor.jinr.ru}}
\setcounter{footnote}{0}
\from{$^{a}$\,Bogoliubov Laboratory of Theoretical Physics, Joint Institute for Nuclear
Research, 6 Joliot-Curie, Dubna, Moscow Region 141980, Russian
Federation}
\from{$^{b}$\,Dubna State University, 19 Universitetskaya, Dubna, Moscow Region
141980, Russian Federation}

\begin{abstract}
Sympathetic cooling of a Fermi gas with a buffer gas of bosonic atoms is an efficient
way to achieve quantum degeneracy in Fermi systems. However, all attempts to use
this method for cooling ions until recently were ineffective because of the unremovable
ion ``micromotion'' in electromagnetic Paul traps, which prevents the realization of a
number of hot projects with cold atom-ion systems. In this regard, we propose a new
efficient method for sympathetic cooling of ions: the use for this purpose of cold buffer
atoms in the region of atom-ion confinement-induced resonances (CIRs) [V.S. Melezhik, Phys. Rev. A103, 53109 (2021)]. We show that the destructive effect of ``micromotion'' on its sympathetic cooling can, however, be suppressed in the vicinity of the atom-ion CIR. Here, the resonant blocking of a close collision of an atom with an ion also resists its heating due to ``micromotion''. We investigate the effect of sympathetic cooling around CIRs in atom-ion, and atom-atom confined collisions within the quantum-quasiclassical approach using the Li-Yb$^+$ and Li-Yb confined systems as an example. In this approach, the Schr\"odinger equation for a cold light atom is integrated simultaneously with the classical Hamilton equations for a hotter heavy-ion or atom during a collision. We have found the region near the atom-ion CIR where the sympathetic cooling of the ion by cold atoms is possible in a hybrid atom-ion trap. We also show that it is possible to improve the efficiency of sympathetic cooling in atomic traps by using atomic CIRs.
\end{abstract}
\vspace*{6pt}


\label{sec:intro}
\section*{Introduction}
The impressive advances in atomic cooling are well known, where temperatures of the order of several nK for alkali metal atoms have been achieved~\cite{Onofrio}. However, it has not yet been possible to reach the same temperature level for cold ions. The best result here is the sympathetic cooling of the Yb$^+$ ions to a temperature of 200~$\mu$K with the help of a cold buffer gas of Li atoms~\cite{Feldker}. This constrains the implementation of hot proposals with cold ions, in particular in the field of metrology and quantum information processing~\cite{Tomza}. Long-term but unsuccessful attempts to achieve a quantum regime with sympathetic cooling of ions nevertheless revealed the main obstacle on this path -- the unremovable ion micromotion caused by the time-dependent radio frequency (RF) fields of the Paul traps used for confining ions~\cite{Grier, Meir, Vuletic, Tomza}. It was shown that the micromotion of the ion and the long-range nature of its interaction with the environment of colder atoms in a hybrid atomic-ion trap prevent the desired effect of sympathetic cooling of ions~\cite{Vuletic, Tomza}. Despite the advances made in sympathetic cooling of ions in hybrid atomic-ion systems in the millikelvin range and above, the problem of cooling to lower energies in these systems is still not resolved.

In our work~\cite{melezhik}, we propose a new way for sympathetic cooling of ions in an electromagnetic Paul trap: to apply for this purpose  buffer cold atoms in the region of the atom-ion confinement-induced resonance (CIR).
Atom-ion CIRs were predicted in ~\cite{MelNegr,Melezhik2019}. It was shown that the CIR occurs when the ratio of the transverse width of the atomic trap $a_{\perp}$ and the s-wave atom-ion scattering length in free space $a_s$ coincides with the value  $a_{\perp}/a_s=1.46$. Earlier, this condition was predicted~\cite{Olshanii} and subsequently confirmed in experiment~\cite{Haller2010} for atomic Cs waveguide-like traps. To describe the dynamics of a quantum particle near CIR, the 1D Fermi pseudopotential with an effective coupling constant $g_{1D}(a_{\perp}/a_s)$ proposed in ~\cite{Olshanii} is successfully used. Atomic CIRs aroused great interest and stimulated research in this direction due to the possibility of using such resonances to tune effective interatomic interactions in a wide range -- from super strong attraction $g_{1D}\rightarrow -\infty$ to super strong repulsion $g_{1D}\rightarrow +\infty$~\cite{Haller2010, Frolich, Haller2009, Selim}. It is also known that at the point $a_{\perp}/a_s$ of CIR the divergence of coupling constant $g_{1D}(a_{\perp}/a_s)$ (and the total reflection) leads to ``fermionization'' of the relative wave-function of the colliding pair whose square modulus behaves the same as for two noninteracting identical fermions~\cite{Olshanii, Selim}. We use the term ``fermionization'' for description of the atom-ion relative dynamics in the atom-ion CIR in the sense of the definition given in \cite{Selim} for the pair of distinguishable particles: in CIR the complete approach of particles one to another is blocked in pair collision, what makes its dynamics similar to the pair collision of identical nontinteracting fermions. This can lead to some compensation of the long-range character of the atom-ion interaction and, as a consequence, to suppression of the micromotion-induced heating during collisions confined by the atom-ion trap. In~\cite{melezhik} it was investigated how the ``fermionization'' near CIR can ``truncate'' the effective atom-ion interaction and suppress the negative effect of the ion micromotion on the cooling of ions. Here, we discuss the principal features of the suggested mechanism and its possible application for sympathetic cooling of ions in hybrid atom-ion traps and for improvement of sympathetic cooling in atomic traps.

\vspace{-0.25\baselineskip}

\label{sec:problem}
\section*{Problem and method}

\begin{figure}
	\centering\includegraphics[width=0.75\textwidth]{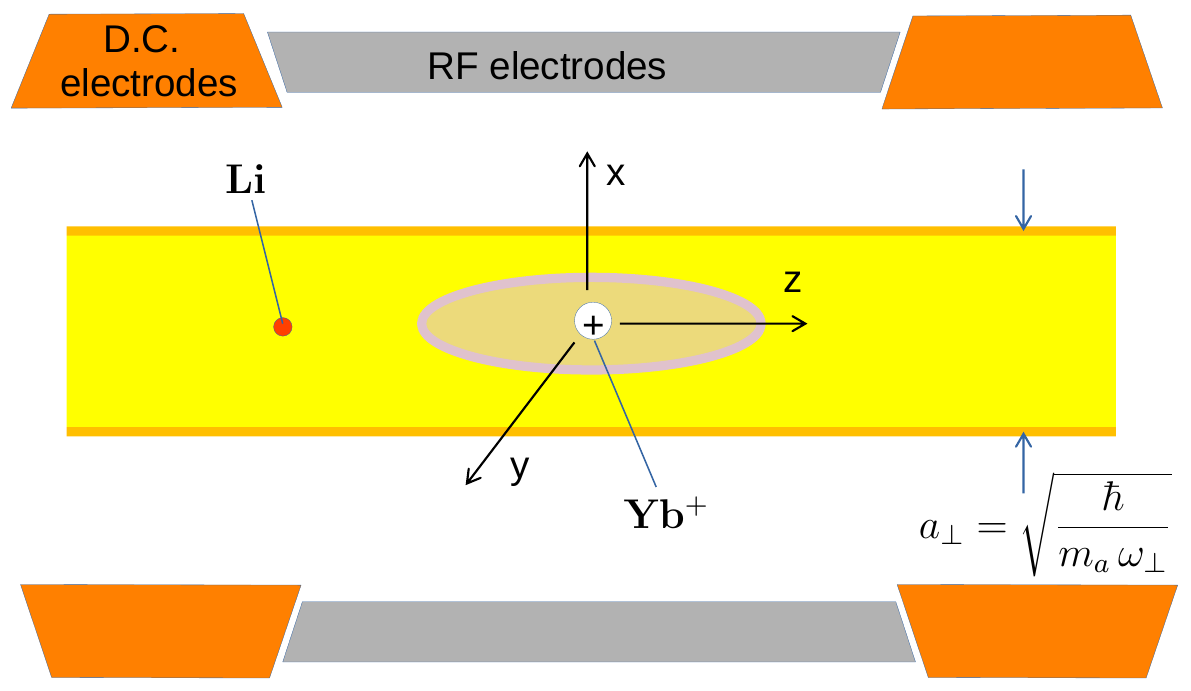}

	\caption{Schematic representation of the atom-ion system confined in a hybrid trap. Here, the ion  is situated in the cloud of cold atoms confined by an optical atomic trap inside the electromagnetic Paul trap. The time-dependent RF field confines the ion transversally, whereas longitudinally a static confinement is formed by the DC field. The dimensions of the confinement region of the ion are determined by the frequencies of the Paul trap $\omega_i$ and $\Omega_{rf}$. The atomic waveguide along the longitudinal axis, $z$, of the linear Paul trap confines the atoms in the transverse $x,y$ directions. The width of the atomic trap $a_{\perp}=\sqrt{\hbar/(m_a\omega_{\perp})}$ is determined by the frequency $\omega_{\perp}$ of the harmonic approximation for the trap shape. Inside the hybrid trap occur paired atom-ion collisions.}
	\label{fig:sketch}
\end{figure}

In ~\cite{melezhik} the effect of sympathetic cooling around CIR in atom-ion confined collisions was investigated within the quantum-quasiclassical method~\cite{MelSchm, Melezhik2001, MelezhikCohen, MelSev, Melezhik2019} using the $^6$Li-$^{171}$Yb$^+$ pair in the hybrid atom-ion trap as an example, which is currently under intense experimental investigations~\cite{JogerPRA17, FuertsPRA18, Feldker}. This specific atom-ion pair is most perspective for sympathetic cooling and reaching the s-wave quantum regime of ions in Paul traps~\cite{Vuletic, Feldker}. The following problem is considered: a hot ion confined in a time-dependent RF Paul trap with linear geometry collides with the cold atom constrained to move into a quasi-one-dimensional waveguide within the ion trap (see Fig.~\ref{fig:sketch}). In our approach~\cite{Melezhik2019, Mel2019}, the Schr\"odinger equation for a cold light atom is integrated simultaneously with the classical Hamilton equations for hotter heavy ion during collision. In the same approach we have also simulated the cooling of hotter heavy atom by cold light atom.

In the developed approach, we calculated the scattering amplitude\linebreak $f(a_{\perp}/a_s)$ of Li atom by an Yb$^+$ ion in the confined geometry of an atomic-ion trap (see Fig.~\ref{fig:sketch}), as well as the mean ion energy
\begin{align}
\label{eq:energii}
\langle E_i^{(\mathrm{out})}\rangle = \frac{1}{t_{\max}-t_{\mathrm{out}}}\int_{t_{\mathrm{out}}}^{t_{\max}}E_i(t)\,dt \,,
\end{align}
after the collision for different initial ion mean energies
$$
\langle E_i^{(\mathrm{in})}\rangle = \frac{1}{t_{\mathrm{in}}}\int_{0}^{t_{\mathrm{in}}}E_i(t)\,dt
$$
as a function of ratio $a_{\perp}/a_s$. When calculating, the times $t_{\mathrm{in}}$ and $t_{\mathrm{out}}$ were chosen in time domains where the atom-ion interaction does not yet effect and no longer affects the ion trajectory, respectively.
In parallel, the effective coupling constant
\begin{align}
\label{eq:g1D}
g_{1D}=\mathop{\lim}\limits_{k
\rightarrow 0}
\frac{\hbar^2 k}{m_a}\frac{\mathrm{Re}[f^{+}(k)]}{\mathrm{Im}[f^{+}(k)]}\,,
\end{align}
which diverges $g_{1D}\rightarrow \pm \infty$ in the CIR region where $a_{\perp}/a_s\rightarrow 1.46$ ~\cite{Olshanii, MelNegr, Melezhik2019}, was calculated from the scattering amplitude. Here, $k =\sqrt{2m_a E_a}/\hbar$, $E_a$, and $m_a$ define the atomic kinetic energy and mass, respectively.

\vspace*{-0.5\baselineskip}

\section*{Results and discussions}\label{sec:results}

\vspace*{-0.5\baselineskip}

\begin{figure}[p]
	\begin{center}
\vspace{-28mm}
		\includegraphics[width=70mm]{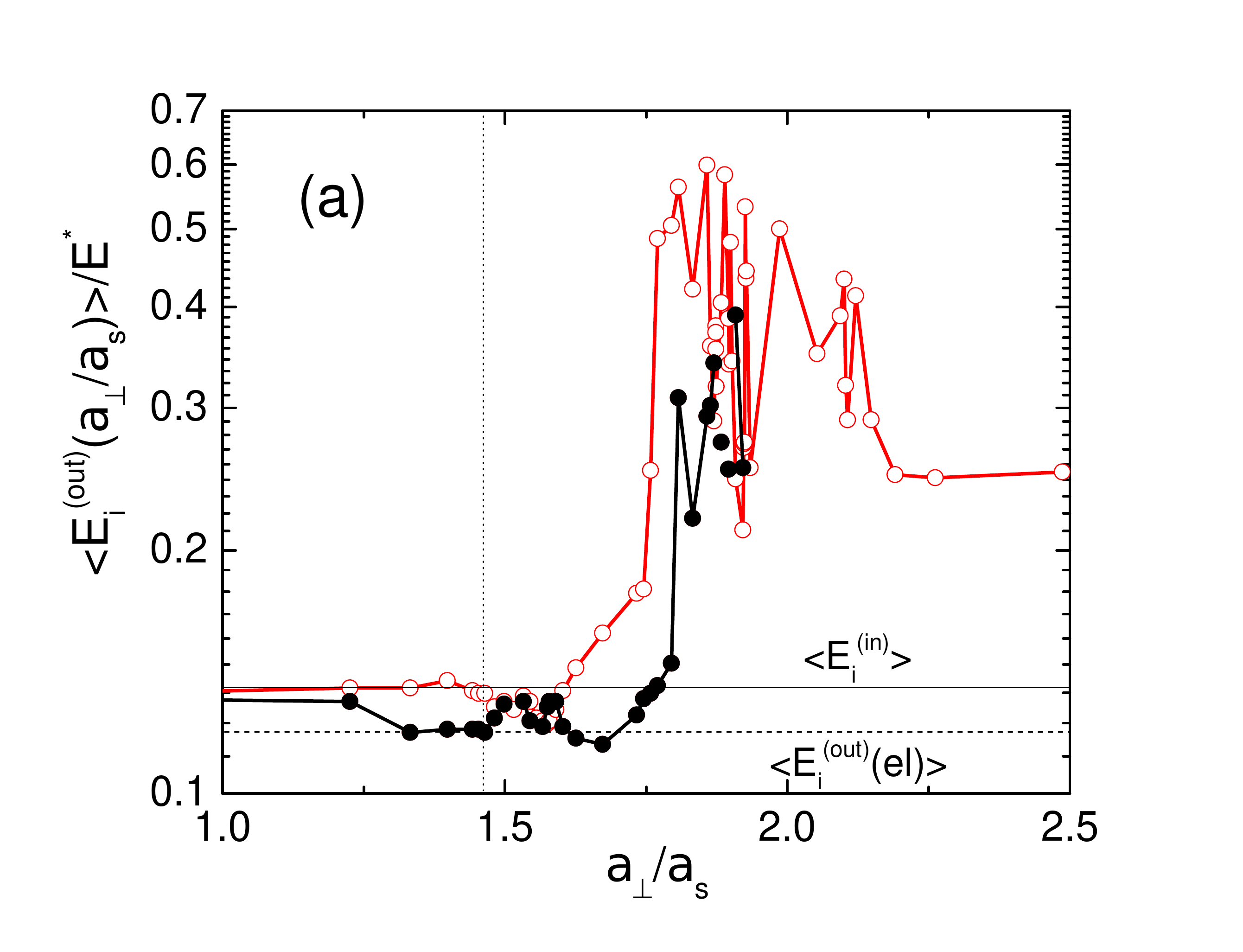}
	\end{center}
	\begin{center}
	\includegraphics[width=70mm]{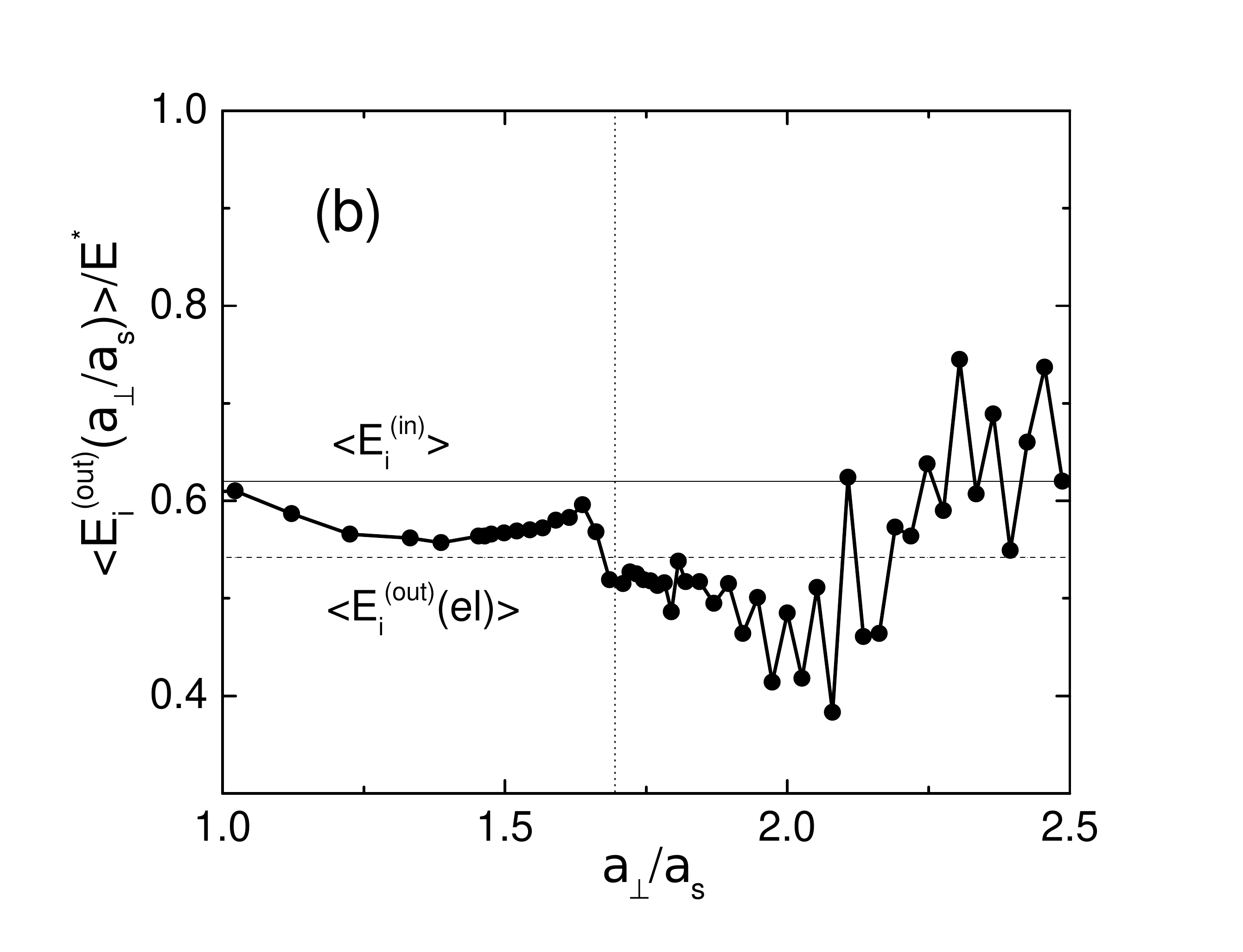}
	\end{center}
	\begin{center}
		\includegraphics[width=70mm]{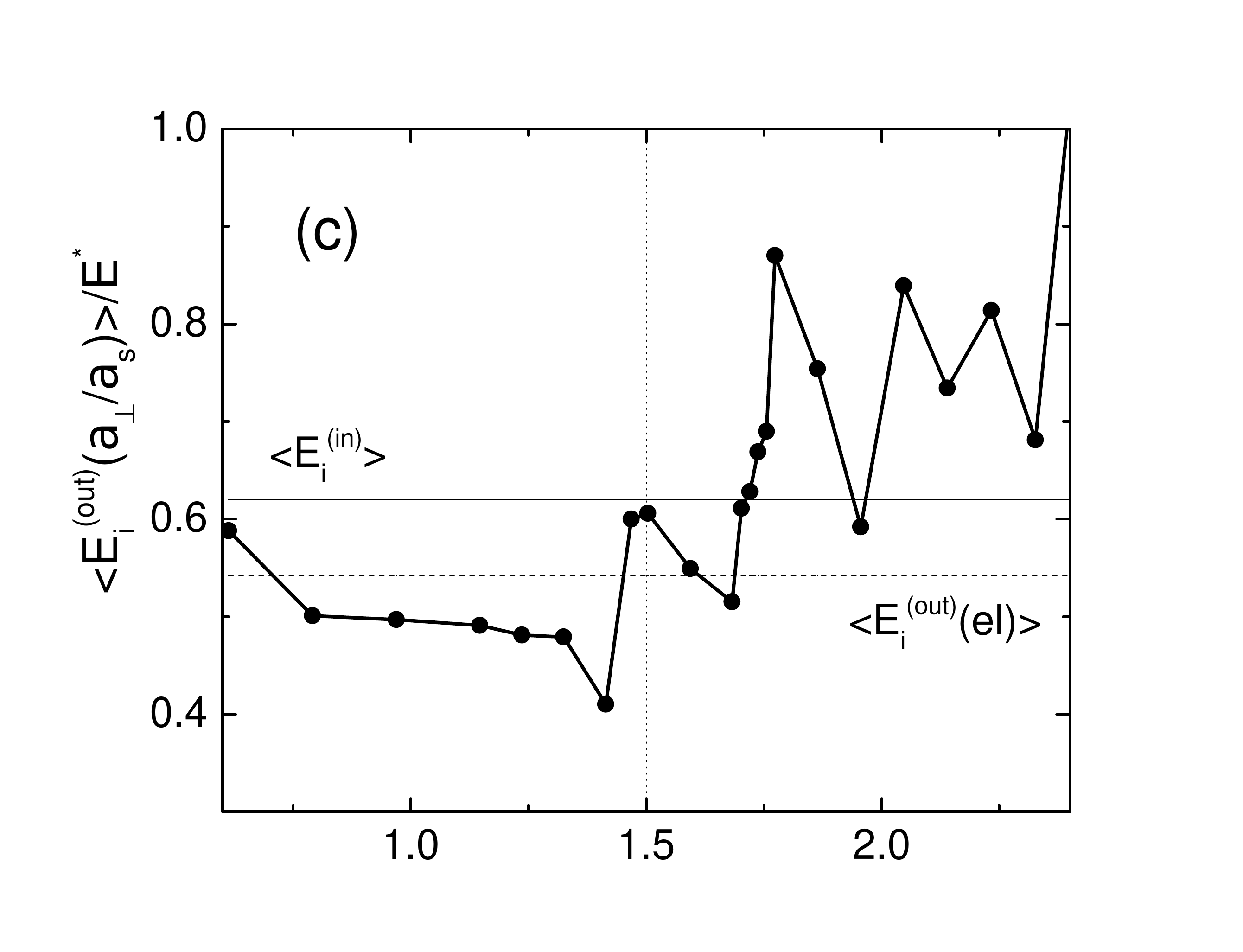}
\caption{(a) The calculated mean ion energy $\langle E^{(\mathrm{out})}_{i}\rangle$ after Li-Yb$^+$ collision. (b) The calculated in the time-independent secular approximation mean ion energy $\langle E^{(\mathrm{out})}_{i}\rangle$ after ion-atom collision. (c) The results of calculations with the potential simulating Li-Yb collision in the time-independent secular approximation. The black circles indicate the results of calculations performed for the ``not head-on collisions'', open circles related to ``head-on collisions''. Here, $E^*=\hbar/(2m_a R^*)$, $R^*=\sqrt{2m_aC_4}/\hbar$, and $C_4$ is the polarizability coefficient for ion-atom pair.} 	
\label{fig02}
\end{center}
\end{figure}

In Fig.~\ref{fig02}a we present calculated dependence of the mean kinetic energy $\langle E^{(\mathrm{out})}_{i}\rangle$  of Yb$^+$ ion after collision with cold Li atom as a function of the ratio $a_{\perp}/a_s$ for the time-dependent RF Paul trap~\cite{Feldker, Tomza, melezhik}. Here we have considered two fundamentally different cases which however correspond to the same mean initial energy of the ion $\langle E_i^{(\mathrm{in})} \rangle$. The first case: in the initial state, the ion has only one transverse momentum component which leads to a ``head-on collisions'' of an ion oscillating in one xz plane with an incident atom moving along the $z$-axis (see Fig.~\ref{fig:sketch}). The second more general case: the ion in the Paul trap before the collision performs 3D motion, what leads to ``not head-on collisions''. One can see that the calculated curves $\langle E_i^{(\mathrm{out})}(a_{\perp}/a_s)\rangle$ have the region of minimal values around the point of the CIR where $a_{\perp}/a_s=1.47$. In Fig.~\ref{fig02}a, the value for the final ion energy $\langle E_i^{(\mathrm{out})}(el)\rangle =\langle E_i^{(\mathrm{in})}\rangle-\Delta E_i(el)$, calculated according to the classical formula
\begin{align}
\label{eq:ellastic}
\Delta E_i(el)= \frac{4m_a m_i}{(m_a+m_i)^2}(\langle E_i^{(\mathrm{in})}\rangle-E_{\mathrm{coll}}-E_{\perp})\,
\end{align}
and valid for the central elastic collision of two classical particles, is also presented. The performed investigation demonstrates
the existence of rather broad region $1.3 \lesssim a_{\perp}/a_s \lesssim 1.7$ where the ion lost energy during ``not head-on collisions'' with the atom approaches the value following from the classical consideration (\ref{eq:ellastic}). This region, however, narrows significantly in the case of  ``head-on collisions''.

Thus, the performed analysis demonstrates that one can control the sympathetic cooling of ions in a hybrid atom-ion trap by tuning the ratio $a_{\perp}/a_s$ to the resonant region around the CIR, where cooling can reach the most optimal conditions. In the cases considered, the maximum loss of energy by an ion reaches the energy corresponding to the loss of energy by a heavy ball with a mass of $^{171}$Yb$^+$ in its elastic head-on collision with a light slow ball with a mass of $^6$Li.

%

We also investigated the effect of a CIR on the sympathetic cooling in a hybrid atomic-ion trap in the framework of the time-independent secular approximation~\cite{LeibfriedRMP03,Melezhik2019} with constant frequencies for the ion-trap interaction. Since the secular approximation for the potential of interaction of an ion with a trap does not contain the RF part and, therefore, does not cause micromotion of an ion, the calculation within its framework allows one to estimate the effect of micromotion-induced  heating at the collision with a cold atom by comparing  with the result in Fig.~\ref{fig02}a including the micromotion. The results of the calculation in the secular approximation are presented in Fig.~\ref{fig02}b where the CIR shifts to the point $a_{\perp}/a_s=1.71$. The demonstrated in Fig.~\ref{fig02}a effect of increasing the sympathetic cooling of an ion in the CIR region is also preserved for the secular approximation in Fig.~\ref{fig02}b. Moreover, this region noticeably expands to the left and right of the CIR point due to absence of the micromotion-induced heating in this case.

Finally, we also evaluated the effect of the long-range character of the atom-ion interaction on the process of sympathetic cooling by replacement of the long-range tail in the interparticle interaction by more short-range Van-der-Waals potential. Results given in Fig.~\ref{fig02}c show that this replacement, which simulates the atom-atom interaction, does not fundamentally change the picture of enhanced sympathetic cooling of a heavy hot particle near CIR. Moreover, in this case we get a deeper minimum for the curve $\langle E_i^{(\mathrm{out})}(a_{\perp}/a_s)\rangle$ to the left of CIR, which is much deeper than the limit following from the model of absolutely elastic central collision of two balls. This consideration can be regarded as a rough model of the quasi-1D Li-Yb scattering in an optical trap in the vicinity of the atomic CIR, which demonstrates that the effect of enhancing the sympathetic cooling of the $^+$Yb-ion in confined collisions with Li-atoms in the resonant region also remains in the confined collision of atomic Yb with cold Li-atoms.

\vspace*{-0.5\baselineskip}

\section*{Conclusion}\label{sec:conclusion}

\vspace*{-0.5\baselineskip}

Based on the performed analysis~\cite{melezhik} we propose a new way for sympathetic cooling of ions in an electromagnetic Paul trap: to use for this purpose cold buffer atoms in the region of atom-ion confinement-induced resonance (CIR). It was also shown that it is possible to improve the efficiency of sympathetic cooling in atomic traps by using atomic CIRs.

In conclusion, atom-ion CIRs have not been experimentally discovered yet. Therefore, it seems to us that it is easier to varify experimentally the predicted mechanism of sympathetic cooling in atomic systems, where magnetic Feshbach resonances are successfully used to tune atomic CIRs~\cite{Haller2010}. Nevertheless, the implementation and tuning of the effective atomic-ion interaction on the atom-ion CIRs may turn out to be a simpler experimental problem than in pure atomic confined systems that does not require the use of the Feshbach magnetic resonance technique. Indeed, the nonresonant s-wave atomic scattering lengths $a_s$ in free space are much smaller than the transverse dimensions of the existing atomic optical traps $a_{\perp}$, that is, in the nonresonant case  $a_{\perp} \gg |a_s|$. Therefore, the realization of the resonance condition $a_{\perp} = 1.46\,a_s$ for atomic systems required the use of the technique of magnetic Feshbach resonances for a sharp increase in the value of $a_s$. In atom-ion systems, the scattering lengths, even in the nonresonant case, significantly exceed the characteristic nonresonant atomic scattering lengths due to the long-range nature of atomic-ion interactions compared to atomic interactions, which leads to the fulfillment of the relation $ a_{\perp} \sim |a_s| $ for atom-ion systems already in the nonresonant case. Therefore, the necessary fine tuning of the atomic-ion scattering length $a_s$ to satisfy the resonance condition $a_{\perp}= 1.46a_s$ can here be replaced by just a slight variation of the atomic trap width $a_{\perp} $ without using the Feshbach resonance technique for enhancement of $a_s$. In this context, the task of experimental detection of the atom-ion CIR and its use for sympathetic cooling of ions in a hybrid atom-ion trap seems to us  quite feasible and relevant.

We also find it interesting to investigate the possibility to use the proposed mechanism for cooling trapped antihydrogen atoms with cold buffer ions.

Author thanks participants of the Conference FFK-2021, P.~Indelicato, S.~Guellati-Kh\'elifa, J.-Ph.~Karr, and V.\,I.~Korobov, for fruitful discussions and suggestions. The work was supported by the Russian Foundation for Basis Research, Grants No.~18-02-00673 and No.~19-02-00058.

\end{document}